\documentclass[reprint,superscriptaddress,showpacs,amsmath,amssymb,aps,prl]{revtex4-1}
\usepackage{graphicx}  
\usepackage{dcolumn}   
\usepackage{bm}        
\usepackage{amssymb}   
\usepackage{mathrsfs} 

\usepackage[colorlinks=True,urlcolor=blue,linkcolor=blue,citecolor=blue]{hyperref}
\usepackage{xcolor}
\usepackage{multirow}
\usepackage{setspace}
\usepackage{CJKutf8}
\usepackage{color,soul}

\usepackage[utf8]{inputenc}
\usepackage[T1]{fontenc}
\usepackage{lmodern}
\usepackage[toc,page]{appendix}
\begin{document}


\begin{CJK*}{UTF8}{}


\title{Hunting Down Magnetic Monopoles in 2D Topological Insulators}

\date{\today}

\author{Wenjie Xi (\CJKfamily{gbsn}奚文杰)}
\affiliation{Tsung-Dao Lee Institute \& School of Physics and Astronomy, Shanghai Jiao Tong University, Shanghai 200240, China}

\author{Wei Ku (\CJKfamily{bsmi}顧威)}
\altaffiliation{corresponding email: weiku@mailaps.org}
\affiliation{Tsung-Dao Lee Institute \& School of Physics and Astronomy, Shanghai Jiao Tong University, Shanghai 200240, China}
\affiliation{Key Laboratory of Artificial Structures and Quantum Control (Ministry of Education), Shanghai 200240, China}


\begin{abstract}
In 2D topological insulators, it is inuitive to imagine a magnetic monopole inside the compact reciprocal space, carrying a quantized charge corresponding to the topological invariant.
However, a theoretical formulation of such a physically appealing picture has proven challenging.
Here, we suggest a simple and useful realization of magnetic monopoles in 2D topological insulators via analytically continuing the system to a third imaginary momentum space.
We then illustrate the evolution of the magnetic monopoles across the topological phase transition and use it to provide natural explanations on: 1) discontinuous jump of integer topological invariants, 2) the semi-metallicity on the phase boundary, and 3) how a change of \textit{global} topology can be induced via a \textit{local} change in reciprocal space.
This generic approach can be applied to studies of all topological states of matter to provide intuitive and convenient understandings alternative to existing considerations.
\end{abstract}
\maketitle
\end{CJK*}

Magnetic monopole is one of the most puzzling particles in the fundamental physics.
It stems from the very original study of electromagnetism~\cite{Dirac1931}, then revives with modern interests of the gauge theory~\cite{Yang1975}, and more recently, grand unified and superstring theories~\cite{Wen1985}.
Contrary to the electric charge that generates the electric field, magnetic charge (namely magnetic monopoles) does not exist in the elementary electromagnetism.
Consequently, magnetic flux lines only form loops and cannot have a source or a sink in nature. 
Thus the duality of electric and magnetic fields is limited to only the cases without sources.
In 1931, P.A.M. Dirac argued that the strict quantized value of elementary electric charge could be explained if there were magnetic monopoles~\cite{Dirac1931}.
After that, in 1975, T.T. Wu and C.N. Yang put forward the appropriate mathematical tool--fiber bundle~\cite{Yang1975}, whose integral formalism links the essential aspects of the properties to the topology of the gauge field, rather than details of the system.
However, although a lot of promising programs are in progress~\cite{Pietila2009,Ray2014}, as a fundamental particle, magnetic monopole is still undiscovered in nature.

\begin{figure}
\includegraphics[width=8.5cm]{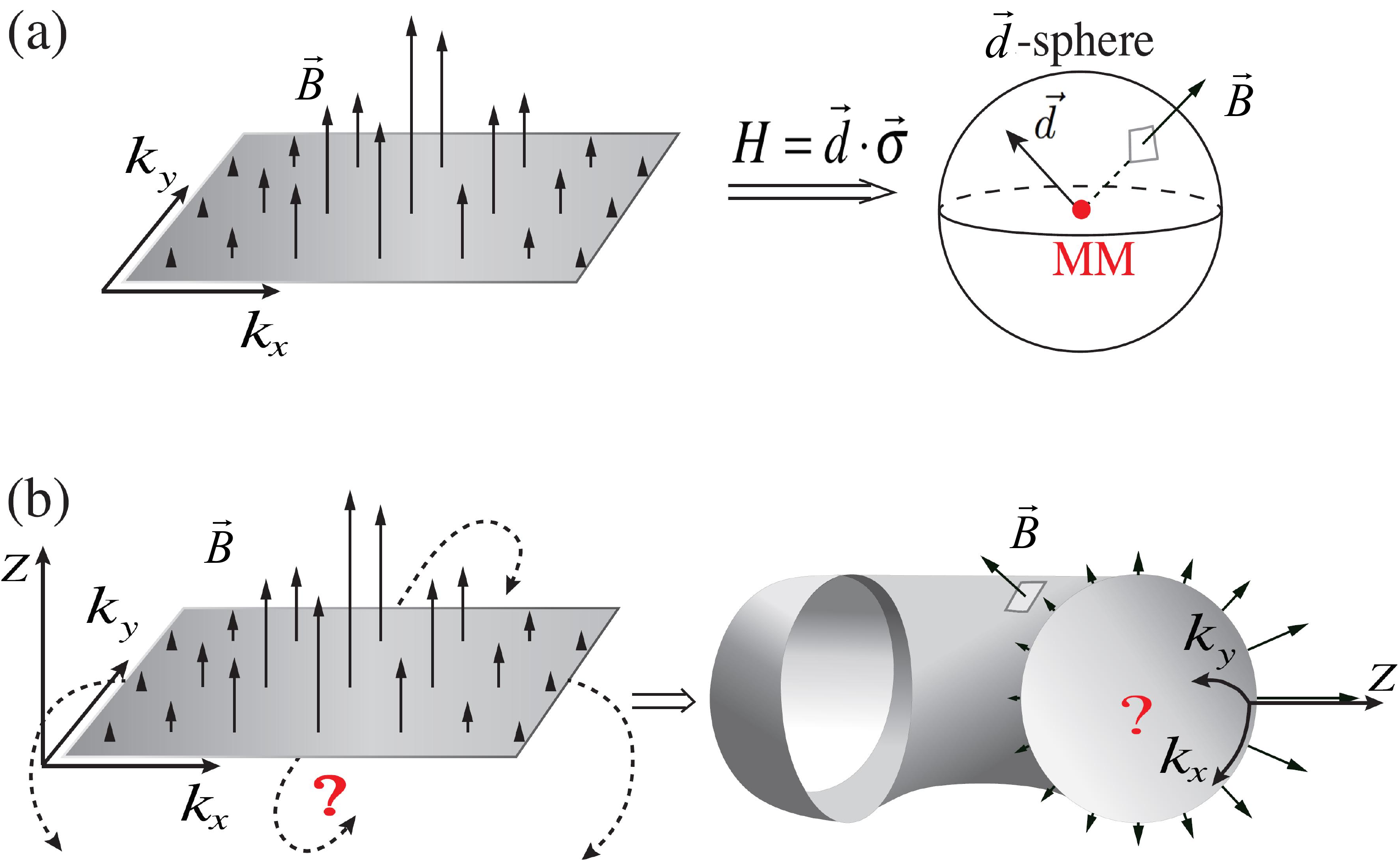}
\caption{\label{fig:fig1}
(color online) Illustration of effective magnetic field in a 2D TI, and simple pictures on how the 2D reciprocal space can be mapped onto (a) a $d$-sphere and (b) a torus via wrapping.
However, in either case, the actual structure of the magnetic monopoles inside the manifold is unknown.
}
\end{figure}

It is thus quite intriguing that magnetic monopoles can be pictured in topological states of matter, such as Quantum Hall effect~\cite{Kohmoto1985, Ku2011}, Topological Insulators (TIs)~\cite{XZ2011, Qi2009, Khom2014}, Weyl semi-metals~\cite{Wan2011,Leon2011,Lu2017}, and spin ice~\cite{Cast2007, Fennell2009, Morris2009, Bramwell2009}.
For example, in 2D TIs, one of the most important class of topological matter, the topological invariant corresponds to the total flux of an effective magnetic field, the Berry curvature, over the entire reciprocal space.
Figure~\ref{fig:fig1}(a) shows the commonly referred picture that once mapped onto the surface of a 3D $d$-sphere, the topological invariant appears to correspond to the total charge of the magnetic monopoles inside the $d$-sphere.
However, other than referring to the long-sought magnetic monopoles, such a physically appealing picture offers little useful information.
For example, through a topological phase transition, this picture offers no explanation how the total charge of the momopole jumps from one integer to another, nor why a semi-metallic state with few emergent Dirac points must exist in the phase boundary.
It certainly does not address one of the most intriguing issue in the TIs:
How can a change of the \textit{gloabl} topology be induced by completely \textit{local} changes in the reciprocal $k$-space near the Dirac points?

Here, we will address this long-standing problem by hunting down the magnetic monopoles in the reciprocal $k$-space of 2D TIs (cf. Fig.~\ref{fig:fig1}(b).)
We will first illustrate the intrinsic difficulty in defining the third dimension and in wrapping the space.
We then demonstrate that analytical continuation to the complex momentum space offers a natural solution in which 1) the magnetic monopoles emerge naturally in pairs at the "branch points" above and below the real axis possessing opposite total charge; and 2) the total charge below the real axis (or inside the torus) gives exactly the topological invariant. 
In essence, the robustness of the topology is mapped to the robustness of the total charge in the lower complex plane, a mapping intriguing even in the mathematical consideration. 
Finally, we will illustrate the evolution of the magnetic monopoles across the topological phase transition and use it to provide natural explanation on 1) discontinuous jump of integer topological invariant, 2) the emergent semi-metallic state at the phase boundary, and 3) how a change of global topology is induced via a local change in reciprocal space.

Let's first demonstrate the intrinsic difficulties of a straightforward approach to choose a third dimension and to wrap the space.
Without loss of generality, consider the well-known BHZ model~\cite{Z2006} expressed by a two-band Hamiltonian in the 2D crystal momentum $k$- and orbital $m$-space:
\begin{eqnarray}
\label{eq:eqn1}
&&\langle\vec{k},m|\hat{H}|\vec{k},m^\prime\rangle\rightarrow\nonumber\\
&&H(\vec{k})=\vec{d}(\vec{k})\cdot \vec{\sigma}=
\left(
\begin{array}{cc}
d_3 & d_1-id_2\\
d_1+id_2 & -d_3
\end{array}
\right),
\end{eqnarray}
where $\vec{d}(\vec{k})=(d_1,d_2,d_3)=(k_x,k_y,M-|\vec{k}|^2/2)$ with a tunable mass perameter $M$.
Expressed in $\vec{b}=\vec{d}/d$,$d=|\vec{d}|$, and a third axis $z$ for simplier 3D notation $\nabla=(\partial_{k_x}, \partial_{k_y}, \partial_z)$, we have the Berry connection (effective magnetic vector potential) $\vec{A}(\vec{k})=i\langle\vec{k},j_0|\nabla|\vec{k},j_0\rangle$ and the Berry curvature (effective magnetic field) $\vec{B}(\vec{k})=\nabla\times\vec{A}=\frac 12 (b_1\nabla b_2\times \nabla b_3+b_2\nabla b_3\times \nabla b_1+b_3\nabla b_1\times \nabla b_2)$, where $|\vec{k},j_0\rangle$ is the eigenvector of the lower band of $H$.
The Chern number~\cite{Thouless1982,Berry1984} (effective magnetic charge)
\begin{eqnarray}
\label{eq:eqn2}
C=\frac{1}{2\pi}\int_{k_x k_y} \vec{B}\cdot \mathrm{d} \vec{S}
\end{eqnarray}
is then zero at negative $M$ (a topological trivial phase) and $C=1$ at positive $M$ (a topological nontrivial phase)~\cite{Z2006}.
Note that owing to the Maxwell theories, the definition of Berry curvature $\vec B$ guarantees that the magnetic charge density
\begin{equation}
\label{eq:eqn5}
\rho=\nabla \cdot \vec B
\end{equation}
is zero anywhere when $\vec A$ is analytical.
(Here for simplicity, the charge are expressed in unit of $1/4\pi$ to give a direct correspondence to their contribution to the Chern number.)
Therefore, the magnetic monopoles can only exist when $\vec A$ and $\vec B$ are non-analytical.

\begin{figure}
\includegraphics[width=8.5cm]{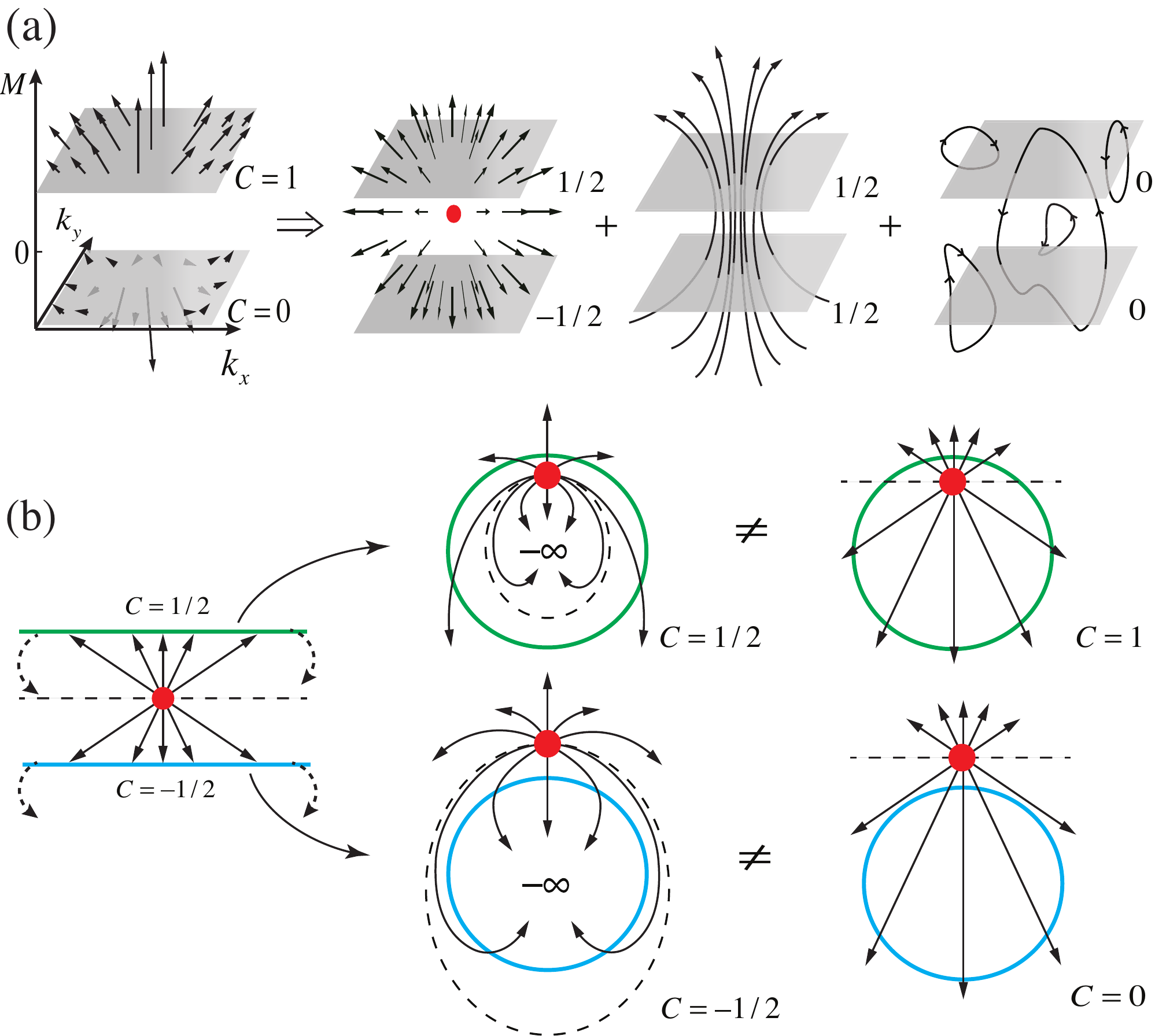}
\vspace{10pt}
\caption{\label{fig:fig2}
(color online) Illustration of the difficulties when wrapping the space in a straightforward construction.
(a) The magnetic monopoles themselves are insufficient to account for the magnetic flux lines.
An additional ``external field'' is necessary to break the symmetry above and below the $z=0$ plane.
(b) The magnetic flux lines will band in the wrapped space, such that only half of the flux lines from charges inside the manifold pass through the enclosed surface, while half of the flux lines from charges outside still contribute to the topological invariant.
}
\end{figure}

Now, to wrap the 2D manifold into a torus and to consider magnetic monopoles inside, one has to define a third dimension.
An easy choice is to pick the $M$ as the $z$ axis.  Since the negative and positive $M$ now have distinct topological invariant, the singular $M=0$ plane must have a topological defect, or a source of magnetic flux.
Indeed, similar to the case of 3D Weyl semi-metals~\cite{Wan2011,Leon2011,Lu2017}, one would find a magnetic momopole of charge 1 at $M=k_x=k_y=0$.
Notice that the introduction of a third dimension in general produces additional $k_x$-$k_y$ components of the effective magnetic field, but they do not change the value of the Chern number on the original physical plane at $M=0$ since only the normal component enters the Chern number (cf: Eq.~\ref{eq:eqn2}).

However, as shown in Fig.~\ref{fig:fig2}(a), since the Chern number $C=0$ below the $M=0$ plane and $C=1$ above, the symmetry of the flux line above and below must be broken by an additional ``external field'' that gives exactly a contribution of 1/2 to the Chern number, in addition to the irrelevant flux loops that do not contribute to the topological invariant.
That is, the magnetic monopoles themselves are insufficient to give the essential physics in such a simple extension.

More serious generic problems occur during banding the space to wrap the physical 2D space (of a specific value of $M$) into a compact manifold.
As shown in Fig.~\ref{fig:fig2}(b), irrespective to the banding the space, only half of the magnetic flux lines generated from a monopole pass through the surface of the torus, giving only $C=1/2$.
For the same reason, a magnetic monopole outside the torus actually gives $C=-1/2$ instead of $C=0$.
This is the basic reason for the necessity of the external field described above, and guarantees the same Chern number, whether the space is wrapped around $M=\infty$ or $M=-\infty$.
But, it also prevents the desired simple picture of just counting the monopoles inside the torus following the Gauss's law in a flat space, shown in the right panels of Fig.~\ref{fig:fig2}(b).

To get around these seemingly inescapable problems, we propose to analytically continue the Hamiltonian to the complex momentum plane, by replacing the original $k_y$ with $k_y+i*z$ along the third dimension $z$.
It has long been known~\cite{Kramers,Kohn1959,Kohn1973,Kohn1974} that in insulators, such an analytical continuation into the complex momentum plane is necessary in order to include the complete set of eigenstates, including the localized states residing inside the band gap (whose decaying amplitudes require complex $k$).
Consequently, the definition of bands should be extended into the band gap, and symmetry-unrelated two-fold degenerate states must exist at the so-called ``branch point'' where the bands from two sides of the gap meet.

In the context of our problem, the well-known eigenstate representation of the $z$-component of the Berry curvature for band $j$~\cite{Thouless1985},
\begin{eqnarray}
\label{eq:eqn3}
B_z^{(j)}(\vec{k})&=&i\sum\limits_{j' \ne j}
\frac{\left\langle {\vec{k},j\left| \partial_{k_x} \widehat{H} \right|\vec{k},j'} \right\rangle \left\langle {\vec{k},j'\left| \partial_{k_y} \widehat{H} \right|\vec{k},j} \right\rangle}{(E_j-E_{j'})^2}\nonumber\\
 &-& \left( {{k_x} \leftrightarrow {k_y}} \right)
\end{eqnarray}
makes it clear that the effective magnetic field must diverge at these branch points due to the zero denominator from the degeneracy.
That is, these branch points mark the location of the magnetic monopoles in this 3D space!

\begin{figure}
\hspace*{-.05in}
\includegraphics[width=8.5cm]{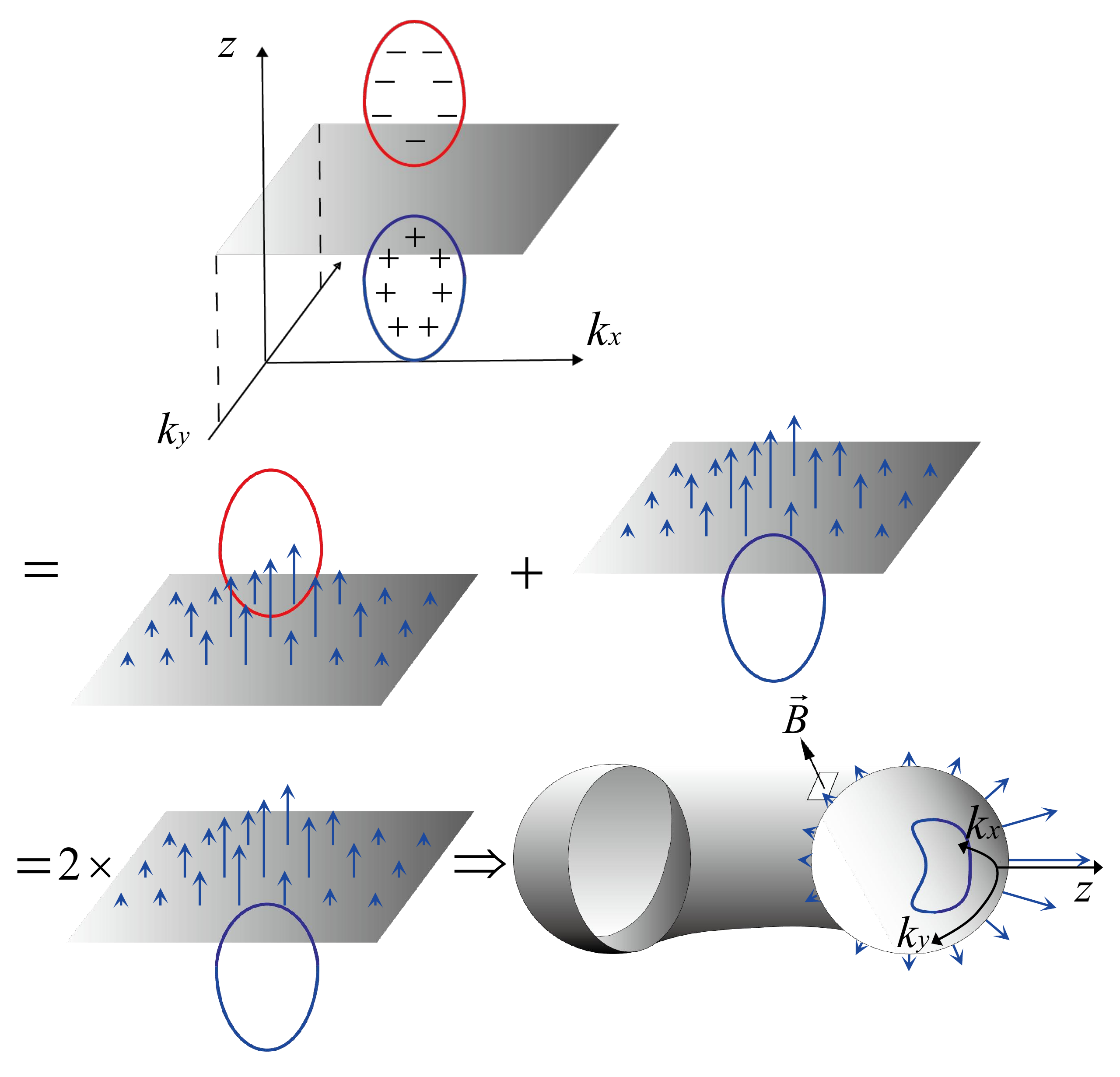}
\vspace{10pt}
\caption{\label{fig:fig3}
(color online) Illustration of structures of magnetic monopoles in our construction.
The magnetic monopoles reside along the strings of branch points, which show clear mirror symmetry in their locations over the $z=0$ plane, but with opposite total charge.
Such symmetry gives exactly the desired feature that the total charge inside the torus corresponds to exactly the topological invariant.
}
\end{figure}

Figure~\ref{fig:fig3} illustrates the structure of the magnetic monopoles in this 3D space for a topological nontrivial case $M=0.2$.
For this specific Hamiltonian, the branch points correspond to a zero eigenenergy $\pm\sqrt{{\vec d} \cdot {\vec d}}$ and form a few closed strings in the 3D space.

One notice immediately that the locations of the monopoles have a mirror symmetry across the $z=0$ plane and the total magnetic charge above and below the plane is exactly opposite.
This generic symmetry provides the foundation for us to workaround the above problems.
As illustrated in Fig.~\ref{fig:fig3}, this symmetry guarantees that the magnetic charge above the $z=0$ plane gives exactly the same contribution to the Chern number as that below.
That is, the total contribution is just twice of that from the negative $z$ space.
At the same time, the infinite $z=0$ plane can only collects half of the total flux from the negative $z$ space, since it only covers a solid angle of $2\pi$ out of the full $4\pi$ angle.
Together, it follows that after wrapping the space around $z=-\infty$, the total flux passing through the torus from all the monopoles is the same as the magnetic charge inside the torus, and introduction of an external field is unnecessary.
Thus, we have successfully built a quantifiable construction of magnetic monopoles to give the desired simple picture, in which the magnetic charge enclosed by the torus gives exactly the topological invariant.

To demonstrate the usefulness of this physically intuitive picture, we investigate below the generic evolution of the magnetic monopoles across the topological phase transition, to provide a simple picture for the following generic features: 1) how the topological invariant jumps from one integer to another, 2) the emergent gapless semi-metallic state, and 3) change of \textit{global} topology through a \textit{local} change in $k$-space.

\begin{figure}
\hspace*{-.35in}
\includegraphics[width=8.5cm]{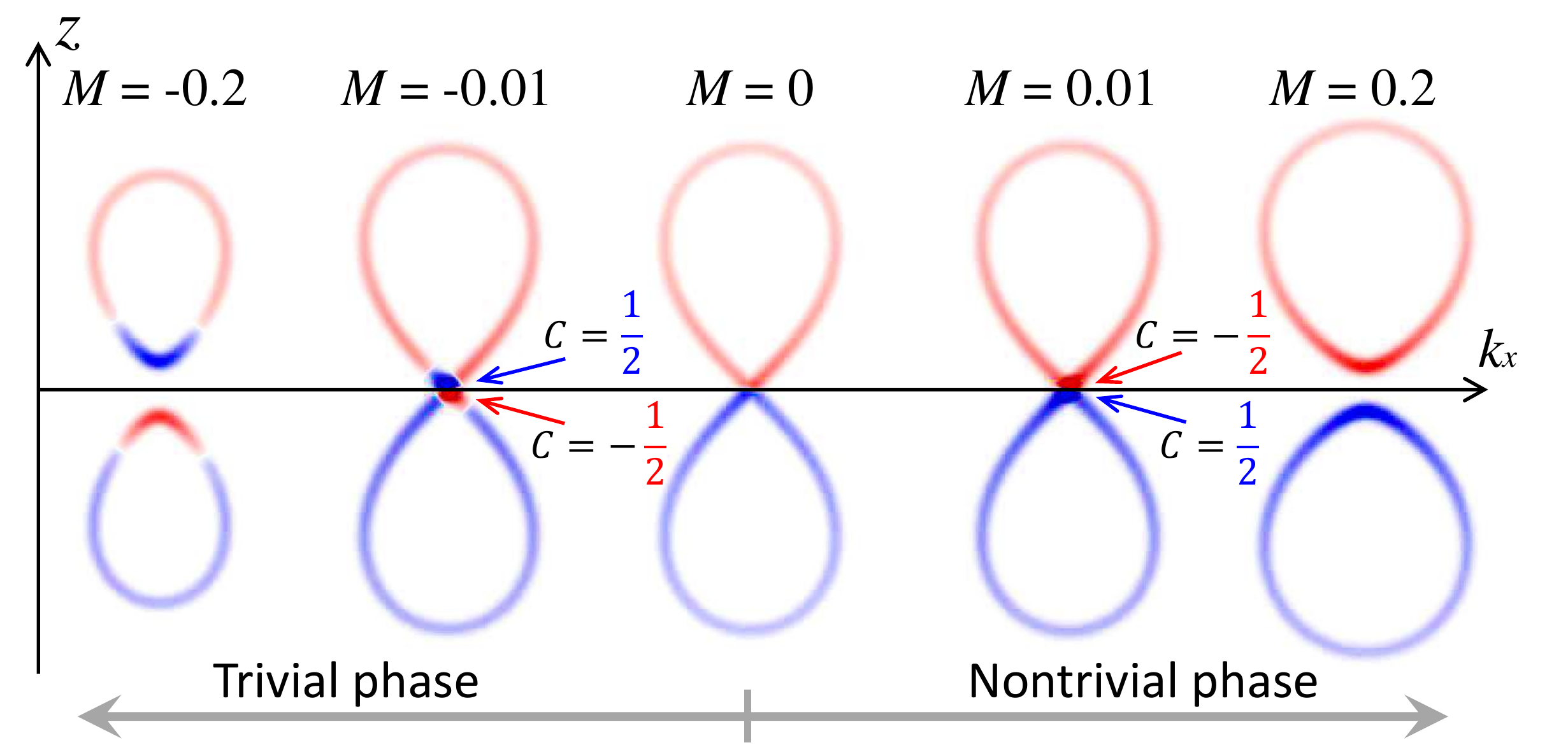}
\vspace{10pt}
\caption{\label{fig:fig4}
(color online) Illustration of evolution of magnetic monopoles across the topological phase transition, from trivial ($M<0$) to nontrivial ($M>0$) phases.
Right before the phase transition, $\pm 1/2$ of the magnetic charges condense into singular points at the tips.
Across the phase transition, the singular $\pm 1/2$ charges move to the other side and thus produce a change of 1 in the topological invariant.
See text for more discussions.
}
\end{figure}

Fig.~\ref{fig:fig4} shows the magnetic monopoles of the BHZ model, from a topological trivial phase $M=-0.2$ to a nontrivial phase $M=0.2$, through the phase boundary at $M=0$.
In the topological trivial phase, the magnetic charge below the $z=0$ plane (those inside the torus after wrapping) contains both positive and negative contributions that sum to a total zero charge, corresponding to a zero Chern number.

As the system approaches the topological phase transition, the tip of the string starts to move closer to the $z=0$ plane, and the negative magnetic charge in red starts to accumulate toward the tip.
Right before the phase transition (c.f. $M=-0.01$), exactly $C=-1/2$ amount of magnetic charge condenses into a singular point at the tip located at $z=0^-$, and a +1/2 amount of charge on the other side $z=0^+$.
(Of course, this implies that the remaining positive contribution in blue below $z=0$ plane integrates to exactly 1/2.)
Right after the phase transition (c.f. $M=0.01$), one find these two singular point charges swap positions: the positive charge moves down to $z=0^-$.
Together with the 1/2 charge spreading out along the string, this gives exactly magnetic charge of size 1, the Chern number of the nontrivial phase.
In other words, the jump of the topological invariant corresponds to swapping condensed 1/2 point charges of opposite sign across the surface of the torus, thus changing the total charge inside by 1.

This analysis also helps to realize the fundamental connection that allows a change of \textit{global} topology via \textit{local} changes in reciprocal space.
By connecting the topological invariant (total magnetic flux through the torus) to the total magnetic charge inside the compact manifold via the Gauss's law, a discontinuous jump in the topological invariant is possible if (and only if) condensed point charges of macroscopic magnitude pass through the surface of the torus, a local phenomenon.
This intriguing connection between global and local properties has profound mathematical implications.
An understanding through residue theorem of the complex analysis will be discussed elsewhere~\cite{He}.

Our construction also gives a physically intuitive picture for the emergent gapless semi-metallic state.
On their way to the other side, this pair of half-charged magnetic monopoles must reside exactly at $z=0$ at the phase boundary $M=0$, as shown in Fig~\ref{fig:fig4}.
At this point, the corresponding eigenstates have purely real momentum $k$, meaning they become regular Bloch states within the standard band structure.
Since these branch points mark the location where the bands above and below the gap meet, the gap must close right at the location of the condensed magnetic monopoles.
In fact, there is a general proportion relationship between $z$ and the gap size, and therefore the linear $z(k)$ dependence at $M=0$ shown in Fig.~\ref{fig:fig4} implies a linear band dispersion near the gap-closed point in this model.
In other words, in our construction, the commonly known ``Weyl'' points or ``Dirac'' points~\cite{Wan2011,Young2012} are nothing but a manifestation of the branch points reaching the real axis carrying a condensed magnetic charge.

It is important to point out that the swapping of condensed magnetic charges is a completely local phenomenon and the corresponding non-analytical ``phase transition'' can be more general than the typical topological phase transitions.
While the topological phase transitions requires such a swapping, the latter does not necessarily lead to a distinct topological phase.
There might exist non-analytical phase transitions that do not involve a change in the topological invariant, for example when such swapping take place at two places with the opposite sign of the swapping.
(In such case, it might be possible to define a new topological invariant to capture the distinct features of the two phases, for example, the spin Chern number~\cite{KM2005,KM2005_2}.)
So, our scheme can actually cover a large scope of non-analytical transition than just the topological insulators.

In summary, we resolve a long-standing issue of quantitatively defining the magnetic monopoles in 2D topological insulators, by analytically continuing the problem onto a complex momentum space.
In such a framework, the magnetic monopoles reside at the branch points that connect the band below and above the gap.
The resulting simple picture has the desired feature that the topological invariant is given by the total charge of the magnetic monopoles inside the 2D compact manifold.
We then show that this physically intuitive picture gives simple descriptions of topological phase transitions: the condensation of pairs of 1/2 magnetic charges of opposite sign, and their passing through the compact manifold.
This description naturally explains the generic features of topological phase transitions:
1) how the topological invariant jumps from one integer to another,
2) the emergent of gapless semi-metallic state at the phase boundary, and
3) how the \textit{global} topology can be induced by a \textit{local} change in the reciprocal space.
Our construction fills an important missing piece of the conceptual picture widely referred to in the literature, and proves to be a useful tool to offer microscopic explanation of almost all the details of bulk electronic structures in topological phases.

We thank Xugang He, Chia-Hui Lin, Genda Gu and Yaomin Dai for very helpful discussions.
Work funded by National Natural Science Foundation of China \#11674220 and 11447601, and Ministry of Science and Technology \#2016YFA0300500 and 2016YFA0300501, and in part by the U. S. Department of Energy,  Office of Basic Energy Sciences DE-AC02-98CH10886.

* weiku@mailaps.org


\begin{thebibliography}{99}

\bibitem{Dirac1931} P.A.M. Dirac, \href{http://rspa.royalsocietypublishing.org/content/133/821/60}{Proc. Roy. Soc. A {\bf{133}}, 60 (1931).} 

\bibitem{Yang1975} T.T. Wu and C.N. Yang, \href{http://journals.aps.org/prd/abstract/10.1103/PhysRevD.12.3845}{Phys. Rev. D \textbf{12}, 3845 (1975).}

\bibitem{Wen1985} X.G. Wen and E. Witten, \href{http://www.sciencedirect.com/science/article/pii/0550321385905929}{Nuclear Physics B{\bf{261}}, 651-677 (1985).}

\bibitem{Pietila2009} V. Pietil$\"a$ and Mikko M$\"o$tt$\"o$nen, \href{http://journals.aps.org/prl/abstract/10.1103/PhysRevLett.103.030401}{Phys. Rev. Lett. {\bf{103}}, 030401 (2009).}



\bibitem{Ray2014} M.W. Ray \emph{et al}., \href{http://www.nature.com/nature/journal/v505/n7485/full/nature12954.html}{Nature {\bf{505}}, 657–660 (2014).}


\bibitem{Kohmoto1985} M. Kohmoto, \href{http://www.sciencedirect.com/science/article/pii/0003491685901484}{ANNALS OF PHYSICS {\bf{160}}, 343-354 (1985).}

\bibitem{Ku2011} P. K. Joshi, P. S. Bisht and O. P. S. Negi, \href{http://www.scirp.org/journal/PaperInformation.aspx?paperID=3636}{JEMAA, {\bf{3}} 1, January (2011).}

\bibitem{Qi2009} X. L. Qi, R. Li, J. Zang and S. C. Zhang, \href{http://science.sciencemag.org/content/323/5918/1184}{Science {\bf{323}} 5918, 1184-1187 (2009).}

\bibitem{XZ2011} X. L. Qi and S. C. Zhang, \href{http://journals.aps.org/rmp/abstract/10.1103/RevModPhys.83.1057}{Review of Modern Physics, \textbf{83}, 1057 (2011).}

\bibitem{Khom2014} D. I. Khomskii, \href{http://www.nature.com/ncomms/2014/140901/ncomms5793/full/ncomms5793.html}{Nature Communications {\bf{5}}, 4793 (2014).}

\bibitem{Wan2011} X. Wan, A. M. Turner, A. Vishwanath and S. Y. Savrasov, \href{http://journals.aps.org/prb/abstract/10.1103/PhysRevB.83.205101}{Phys. Rev. B \textbf{83}, 205101 (2011).}

\bibitem{Leon2011} Burkov, A. A. and Balents, Leon, \href{https://journals.aps.org/prl/abstract/10.1103/PhysRevLett.107.127205}{Phys. Rev. Lett \textbf{107}, 127205 (2011).}


\bibitem{Lu2017} Y.X.Zhao and Y.Lu, \href{https://journals.aps.org/prl/abstract/10.1103/PhysRevLett.118.056401}{Phys. Rev. Lett. {\bf{118}}, 056401 (2017).}



\bibitem{Cast2007} C. Castelnovo, R. Moessner, and S. L. Sondhi, \href{http://www.nature.com/nature/journal/v451/n7174/abs/nature06433.html}{Nature {\bf{451}}, 42–45 (2007).}

\bibitem{Fennell2009} T. Fennell, \emph{et al}., \href{http://science.sciencemag.org/content/326/5951/415}{cience {\bf{326}}, 415–417 (2009).}

\bibitem{Morris2009} D. J. P. Morris, \emph{et al}., \href{http://science.sciencemag.org/content/326/5951/411}{Science {\bf{326}}, 411–414 (2009).}

\bibitem{Bramwell2009} S. T. Bramwell, \emph{et al}., \href{http://www.nature.com/nature/journal/v461/n7266/abs/nature08500.html}{Nature {\bf{461}}, 956–4211 (2009).}




\bibitem{Z2006} B. A. Bernevig, T. L. Hughes, S. C. Zhang, \href{http://www.sciencemag.org/content/314/5806/1757.short}{Science \textbf{314}, 1757-1761 (2006).}

\bibitem{Thouless1982} D. J. Thouless, M. Kohmoto, M. P. Nightingale, and M. Den Nijs, \href{http://journals.aps.org/prl/abstract/10.1103/PhysRevLett.49.405}{Phys. Rev. Lett. \textbf{49}, 405 (1982).}

\bibitem{Berry1984} M.V. Berry, \href{http://rspa.royalsocietypublishing.org/content/392/1802/45}{Proc. R. Soc. Lond. A {\bf{392}}, 45-57 (1984).}




















\bibitem{Kramers} H. A. Kramers, \href{http://www.sciencedirect.com/science/article/pii/S0031891435901185#}{Physica \textbf{2}, 483-490 (1935).} 

\bibitem{Kohn1959} W. Kohn, \href{http://journals.aps.org/pr/abstract/10.1103/PhysRev.115.809}{Phys. Rev. \textbf{115}, 809 (1959).}

\bibitem{Kohn1973} J. J. Rehr and W. Kohn, \href{http://journals.aps.org/prb/abstract/10.1103/PhysRevB.9.1981}{Phys. Rev. B \textbf{9}, 1981 (1973).}

\bibitem{Kohn1974}  J. J. Rehr and W. Kohn, \href{http://journals.aps.org/prb/abstract/10.1103/PhysRevB.10.448}{Phys. Rev. B \textbf{10}, 448 (1974).}






\bibitem{Thouless1985} Niu, Qian and Thouless, D. J. and Wu, Yong-Shi, \href{https://link.aps.org/doi/10.1103/PhysRevB.31.3372}{Phys. Rev. B \textbf{31}, 3372 (1985).}


\bibitem{He} Xu-Gang He and Wei Ku, to be published,




\bibitem{Young2012}Young, S. M. and Zaheer, S. and Teo, J. C. Y. and Kane, C. L. and Mele, E. J. and Rappe, A. M., \href{https://journals.aps.org/prl/abstract/10.1103/PhysRevLett.108.140405}{Phys. Rev. Lett 
\textbf{108}, 140405 (2012).}

\bibitem{KM2005} C. L. Kane and E. J. Mele, \href{http://journals.aps.org/prl/abstract/10.1103/PhysRevLett.95.146802}{Phys. Rev. Lett. \textbf{95}, 146802 (2005).}

\bibitem{KM2005_2} C. L. Kane and E. J. Mele, \href{https://journals.aps.org/prl/abstract/10.1103/PhysRevLett.95.226801}{Phys. Rev. Lett. \textbf{95}, 226801 (2005).}

\end{thebibliography}
\end{document}